\documentclass[aps,preprint,groupedaddress,a4paper]{revtex4-2}
\usepackage{graphicx}

\usepackage{mathtools}
\usepackage[colorlinks = true,linkcolor = blue,urlcolor  = blue,citecolor = blue,anchorcolor = blue]{hyperref}
\usepackage{xcolor, soul}
\usepackage{color}

\usepackage{makecell}

\DeclarePairedDelimiter\ket{\lvert}{\rangle}
\DeclarePairedDelimiter\bra{\langle}{\rvert}

\begin{document}

%\preprint{}

\title{Fault-tolerant qubit encoding using a spin-7/2 qudit}


\author{Sumin Lim}
%\thanks{}
\affiliation{CAESR, Department of Physics, University of Oxford, The Clarendon Laboratory, Parks Road, Oxford OX1 3PU, UK}
\author{Junjie Liu}
%\thanks{}
\affiliation{CAESR, Department of Physics, University of Oxford, The Clarendon Laboratory, Parks Road, Oxford OX1 3PU, UK}
\author{Arzhang Ardavan}
%\thanks{}
\affiliation{CAESR, Department of Physics, University of Oxford, The Clarendon Laboratory, Parks Road, Oxford OX1 3PU, UK}


\date{\today}

\begin{abstract}
The implementation of error correction protocols is a central challenge in the development of practical quantum information technologies. Recently, multi-level quantum resources such as harmonic oscillators and qudits have attracted interest in this context because they offer the possibility of additional Hilbert space dimensions in a spatially compact way. Here we propose a quantum memory, implemented on a spin-7/2 nucleus hyperfine-coupled to an electron spin-1/2 qubit, which provides first order $X$, $Y$ and $Z$ error correction using significantly fewer quantum resources than the equivalently effective qubit-based protocols. Our encoding may be efficiently implemented in existing experimentally realised molecular electron-nuclear quantum spin systems. The strategy can be extended to higher-order error protection on higher-spin nuclei.
\end{abstract}

%\keywords{}

\maketitle

On the path towards a universal quantum computer, there is a broad consensus that we are now in the noisy-intermediate-scale-quantum (NISQ) era~\cite{Preskill2018}. Although some reports~\cite{Arute2019,PhysRevLett.127.180501} suggest that quantum supremacy is possible even with NISQ devices, we are currently limited to certain specific tasks and to computations on the scale of tens of qubits. Progress beyond this will depend on the development of reliable quantum error correction strategies~\cite{RevModPhys.87.307,PhysRevLett.77.3260,PhysRevLett.77.198}, because the major limitation to scalability is the rapid drop in fidelity owing to environmental noise as the system grows. 

A qubit-based quantum error correction algorithm employs  additional physical qubits to encode logical qubits~\cite{PhysRevA.55.900,PhysRevLett.84.2525,GoogleSurfaceCode2023} by providing redundancy in the Hilbert space to protect the information. Although this approach has the advantage of being mathematically compact~\cite{Gottesman-arXiv1997} (which means it is scalable), its implementation in real systems poses challenges arising from the inherent proliferation of quantum resources required. Thus, in the current NISQ era, the traditional error-correction approach offers the dilemma of adding noisy qubits to the quantum system in an attempt to reduce noise.

This highlights the imperative of identifying hardware-efficient implementations in which error correction can be incorporated with minimal quantum resource requirements. In this context error correction algorithms exploiting not only qubits but also qudits (physical systems each offering a $d$-dimensional Hilbert space, with $d>2$ in general) have attracted attention in recent years.

Among the best-known is the Gottesman-Kitaev-Preskill (GKP) code and its expansion~\cite{PhysRevA.64.012310,PhysRevA.77.032309,PhysRevLett.125.080503}. These proposals explore physical systems described by the quantum harmonic oscillator because, in principle, it can provide an infinite-dimensional bosonic Hilbert space for information encoding. Implementations can be provided by systems such as trapped ions or superconducting circuits~\cite{Hu_2019,PhysRevLett.125.043602}. Indeed, some experiments~\cite{Ofek_2016,Campagne_Ibarcq_2020} have already shown enhanced coherence times for quantum states. Various theoretical programmes~\cite{PhysRevX.10.031050,PhysRevLett.125.260509} to generalize bosonic error correction codes are in progress.

A similar but alternative approach is to use a spin qudit, which can provide an intrinsically bounded $d$-level system. Qudits may be realised, for example, by electronic and nuclear spins in the solid state. Although spin-based quantum computing (specifically, exploiting ensembles of molecules whose nuclear spins were driven and detected using nuclear magnetic resonance) attracted attention and yielded some milestone experimental results in the early days of the field ~\cite{vandersypen2001experimental,gershenfeld1997bulk,cory1997ensemble,jones1998implementation} , intrinsic limitations on scalability were quickly identified~\cite{warren1997usefulness}.
However, recent remarkable progress on studies related to molecular magnets encourages us to explore these ingredients as basic building blocks for quantum information processing ~\cite{leuenberger2001quantum,tejada2001magnetic,gaita2019molecular,wasielewski2020exploiting,ardavan2007will}. 
 
The size, number of spins and basic Hamiltonian parameters for molecular magnets can be carefully tuned by chemical engineering, for example, by the selection from various options for ligand cages of a choice of magnetic ions. Furthermore, strategies for single-qudit-molecule addressing are active areas of research. For example, by analogy with the single electron transistor quantum dot, control and readout of a molecular nuclear spin in a single-molecule transistor has been reported ~\cite{vincent2012electronic,thiele2014electrically}. Electrical control of spin \--- which is essential for fast manipulation and spatially-localised control \--- is also being studied by various research groups ~\cite{sigillito2017all,wolfowicz2014conditional,liu2021quantum}. The fact that spin is a basic quantum property of matter which is often only weakly coupled to other degrees of freedom suggests that may be a useful and error-robust embodiment for encoding quantum information.

In this context, we propose a strategy for constructing compact and hardware-efficient information encoding in a single spin qudit. As with the GKP code, it can provide $N^\mathrm{th}$ order error correction in a Hilbert space with dimension of order $N^2$, which is significantly smaller than the $\sim $exp$(N)$ scaling for qubit implementations. 
However, while the GKP code is implemented on a finite subspace of an infinite-dimensional space, this encoding targets a finite-dimensional space, and we find that we need a volume half as big as for GKP.
Previous reports have proposed spin qudit QEC algorithms for correcting  phase errors~\cite{chiesa2020molecular,lockyer2021targeting,carretta2021perspective},
or logical code-words to correct first-order rotation errors using one large spin~\cite{PhysRevLett.127.010504}. In this letter, we adapt a proposed encoding~\cite{PhysRevLett.127.010504} to describe a practical implementation yielding full first-order error correction based on an electron spin qubit and a hyperfine-coupled nuclear spin-7/2, addressing the required hardware-level implementation operations, the gain offered by the protocol, and the required experimental operation fidelities. We offer a generalization to higher-order correction cases, with example codewords.  

Our protocol can be implemented directly in a quantum processor based on nuclear spins in the solid state (including, for example, spin-7/2 nuclei such as $^{51}$V, $^{165}$Ho or $^{123}$Sb)~\cite{atzori2021radiofrequency,liu2021quantum,asaad2020coherent}. Since, among spin-qubit candidates, nuclear spins usually exhibit superior coherence times, they have already been explored as quantum information storage units in many pivotal proposals, such as Kane's early model using phosphorus donors in silicon~\cite{kane1998silicon}, nitrogen vacancy centres in diamond ~\cite{abobeih2018one}, or hybrid structures incorporating electrical 
circuits and molecular magnets ~\cite{moreno2018molecular,thiele2014electrically}.

The protocol relies on projective measurement of the nuclear spin state. From a practical perspective, in the solid state this is typically achieved via measurement of a coupled electronic spin. We therefore base our proposal on an electronic spin-1/2 (acting as the ancillary detection and ``interface'' qubit) which is hyperfine-coupled to a spin-7/2 nucleus (the data storage qudit); this represents a Hilbert space extension of $d=8$ over the qubit space. 

We remark that a quantum information processor unit (based on the $^{123}$Sb donor in Si) has already been realised experimentally that exhibits these properties and all of the necessary operations for implementation of the protocol~\cite{asaad2020coherent}. We would expect our protocol to yield substantially enhanced relaxation and coherence times, thereby offering a highly efficient structure for robust quantum data storage.     

\begin{figure}
\includegraphics{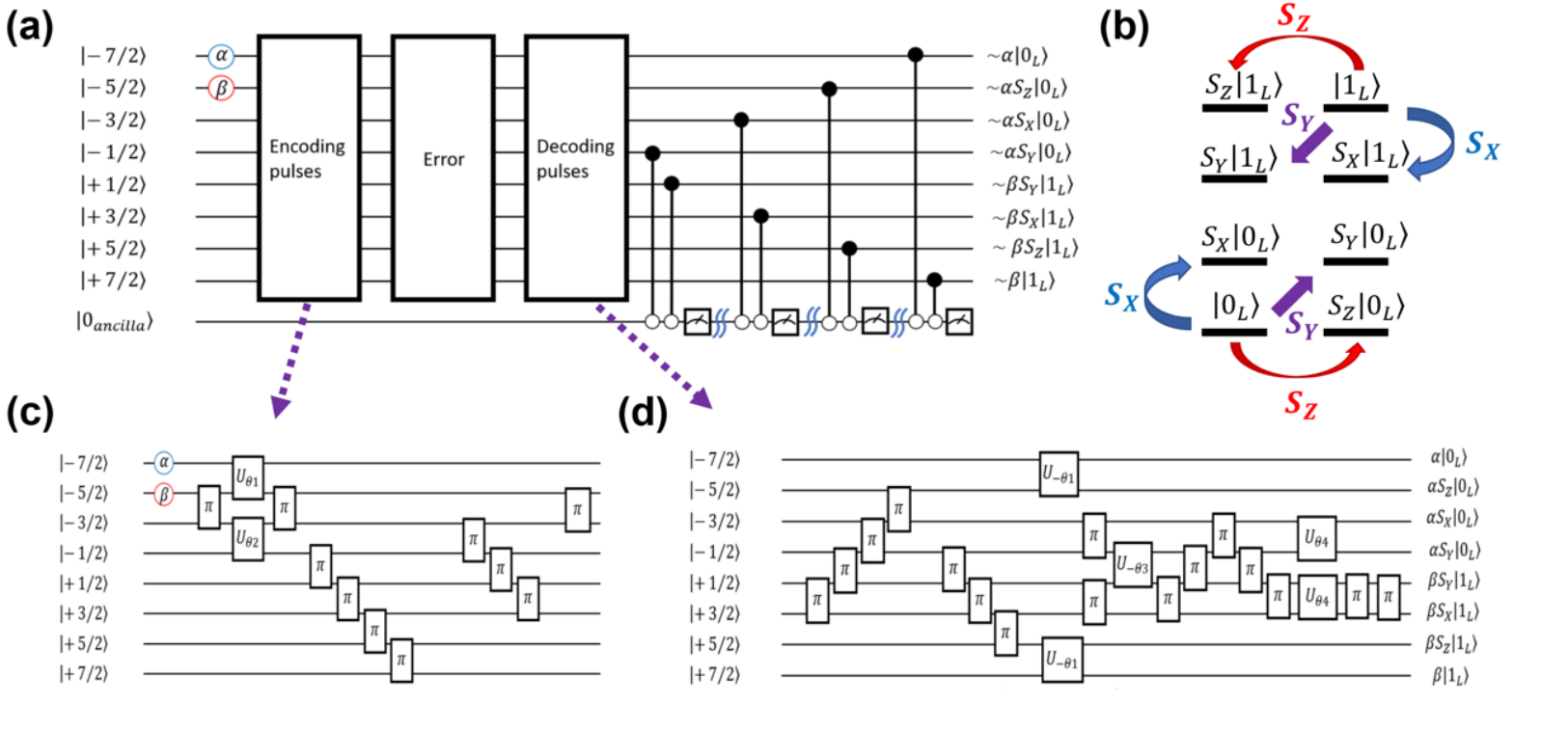}
\caption{A brief schematic of: (a) the entire QC code; (b) the code word with first-order error; (c) the encoding pulse sequence $U_\mathrm{enc}$; and (d) decoding pulse sequence $U_\mathrm{dec}$. Unitary rotations $U_{\theta_i}$ are around the $y$-axis, with $\cos \theta_i = \sqrt{3/10}, \sqrt{7/10}, \sqrt{1/5}, \sqrt{1/2} $ for $i=1,2,3,4$ respectively.
\label{schematic}}
\end{figure}

A schematic diagram is presented in Fig.~ \ref{schematic}. The quantum state to be protected,
$\ket{\psi} = \alpha \ket{0} + \beta \mathrm{e}^{{\mathrm i}\phi} \ket{1} $ 
will be encoded into 
$U_\mathrm{enc} \ket{\psi} = \ket{\Psi^\mathrm{enc}} = \alpha \ket{0_L} + \beta \mathrm{e}^{{\mathrm i}\phi} \ket{1_L} $ 
by the encoding pulse sequence in Fig.~\ref{schematic}(c), with logical code words 
\begin{equation}
\ket{0_L} = \sqrt{\frac{3}{10}} \ket*{-\frac{7}{2}} + \sqrt{\frac{7}{10}}\ket*{+\frac{3}{2}}
\:\:\mathrm{and}\:\:
\ket{1_L} = -\sqrt{\frac{7}{10}} \ket*{-\frac{3}{2}} + \sqrt{\frac{3}{10}}\ket*{+\frac{7}{2}}
\end{equation}
This logical qubit encoding is designed to handle phase ($Z$), shift ($X$) and phase-shift ($Y$) errors simultaneously. The time-evolving  density matrix accumulating errors can be approximately described by~\cite{chiesa2020molecular}
\begin{equation}
\rho(t) = \sum_{k,i} E_{k,i} \rho(0) E_{k,i}^\dagger
\end{equation}
with error operator
\begin{equation}
E_{k,i} \sim \sqrt{ \frac{(t/T_{\mathrm{relax},i})^k}{k!} } S^k_i
\end{equation}
where $i=X,Y$, or $Z$, $T_{\mathrm{relax},i}$ is the typical relaxation time of the system, and $k = 0 \cdots \infty$. We may assume that for times short compared to $T_\mathrm{relax}$, only first order ($i=1$) terms contribute significantly to the error, i.e.,
\begin{equation}
\rho(t) = (1-\epsilon)I 
+ \epsilon_X S_X \rho(0) S_X^\dagger
+ \epsilon_Y S_Y \rho(0) S_Y^\dagger
+ \epsilon_Z S_Z \rho(0) S_Z^\dagger
+ O(\epsilon^2 \mathrm{\, and \, higher}) ,
\end{equation}
where $\epsilon_i \sim t / T_{\mathrm{relax},i}$ is indicative of the scale of the error. We note here that unlike for spin qubits (for which $S_i^2 = I$), the higher-order qudit spin operators $S_i^k$ contribute to higher order errors.

Thus, following application of the error operator, the state becomes corrupted to
\begin{eqnarray}
\ket{\Psi^\mathrm{enc+error}} & = &
\sqrt{1-\epsilon} \left( \alpha \ket{0_L} + \beta \mathrm{e}^{{\mathrm i}\phi} \ket{1_L} \right) + \nonumber \\
& & \sqrt{\epsilon_X} \left( \alpha S_X \ket{0_L} + \beta \mathrm{e}^{{\mathrm i}\phi} S_X \ket{1_L} \right) +  \nonumber \\
& & \sqrt{\epsilon_Y} \left( \alpha S_Y \ket{0_L} + \beta \mathrm{e}^{{\mathrm i}\phi} S_Y \ket{1_L} \right) + \nonumber \\
&  & \sqrt{\epsilon_Z} \left( \alpha S_Z \ket{0_L} + \beta \mathrm{e}^{{\mathrm i}\phi} S_Z \ket{1_L} \right) + \mathrm{\, higher \, order} 
\end{eqnarray}
As required if we are to use these states for quantum error correction, the states in this superposition satisfy the Knill-Laflamme criteria~\cite{PhysRevA.55.900}. Under the action of the error operator the original code words $\ket{0_L}$, $\ket{1_L}$ are transformed to span the states 
$\ket{0_L}$, $\ket{1_L}$, 
$S_X \ket{0_L}$, $S_X \ket{1_L}$,
$S_Y \ket{0_L}$, $S_Y \ket{1_L}$,
$S_Z \ket{0_L}$ and $S_Z \ket{1_L}$, 
(see Table~\ref{enc_space}) and these eight states are mutually orthonormal; they satisfy the conditions
\begin{eqnarray}
\bra{0_L} S_i^l S_j^k \ket{1_L} & = & 0 \:\:\mathrm{and} \nonumber \\
\bra{0_L} S_i^l S_j^k \ket{0_L}   = 
\bra{1_L} S_i^l S_j^k \ket{1_L}  & = & \delta_{ij} \delta_{lk}
\end{eqnarray}
where $i, j = X, Y$ or $Z$ and $l, k = 0$ or $1$. 

\begin{table}[htp]
\caption{Code words and Error-spaces for spin 7/2 encoding}
\begin{center}
\begin{tabular}{|c|c|}
\hline
Logical code word & Representation in $z$-basis \\
\hline\hline
$\ket{0_L}$ & 
$\sqrt{+\frac{3}{10}} \ket*{-\frac{7}{2}} + \sqrt{\frac{7}{10}}\ket*{+\frac{3}{2}}$ \\ \hline
$\ket{1_L}$ & 
$-\sqrt{\frac{7}{10}} \ket*{-\frac{3}{2}} + \sqrt{\frac{3}{10}}\ket*{+\frac{7}{2}}$ \\ \hline
$S_X \ket{0_L} $ & 
$+\sqrt{\frac{1}{10}} \ket*{-\frac{5}{2}} + \sqrt{\frac{1}{2}}\ket*{+\frac{1}{2}} +\sqrt{\frac{2}{10}} \ket*{+\frac{5}{2}} $\\ \hline
$S_X \ket{1_L}$ & 
$-\sqrt{\frac{2}{10}} \ket*{-\frac{5}{2}} - \sqrt{\frac{1}{2}}\ket*{-\frac{1}{2}} +\sqrt{\frac{1}{10}} \ket*{+\frac{5}{2}} $\\ \hline
$\mathrm{i}S_Y \ket{0_L}$ & 
$-\sqrt{\frac{1}{10}} \ket*{-\frac{5}{2}} + \sqrt{\frac{1}{2}}\ket*{+\frac{1}{2}} -\sqrt{\frac{2}{10}} \ket*{+\frac{5}{2}} $\\ \hline
$\mathrm{i}S_Y \ket{1_L}$ & 
$-\sqrt{\frac{2}{10}} \ket*{-\frac{5}{2}} + \sqrt{\frac{1}{2}}\ket*{-\frac{1}{2}} +\sqrt{\frac{1}{10}} \ket*{+\frac{5}{2}} $\\ \hline
$S_Z \ket{0_L}$ & 
$-\sqrt{\frac{7}{10}} \ket*{-\frac{7}{2}} + \sqrt{\frac{3}{10}}\ket*{+\frac{3}{2}}$ \\ \hline
$S_Z \ket{1_L}$ &
$+\sqrt{\frac{3}{10}} \ket*{-\frac{3}{2}} + \sqrt{\frac{7}{10}}\ket*{+\frac{7}{2}}$ \\ 
\hline
\end{tabular}
\end{center}
\label{enc_space}
\end{table}

This implies that first order errors on the logical state can be detected and corrected. Indeed, the sequence of pulses in Fig.\ref{schematic}(d) transforms the corrupted state into 
\begin{eqnarray}
U_\mathrm{dec} \ket{\Psi^\mathrm{enc+error}} & = & 
\sqrt{1-\epsilon} \left( \alpha \ket*{-\frac{7}{2}} +\beta \mathrm{e}^{\mathrm{i}\phi} \ket*{+\frac{7}{2}} \right) + \nonumber \\
& &  \sqrt{\epsilon_Z} \left( \alpha \ket*{-\frac{5}{2}} +\beta \mathrm{e}^{\mathrm{i}\phi} \ket*{+\frac{5}{2}} \right) + \nonumber \\
& & \sqrt{\epsilon_X} \left( \alpha \ket*{-\frac{3}{2}} +\beta \mathrm{e}^{\mathrm{i}\phi} \ket*{+\frac{3}{2}} \right) + \nonumber \\
& & \sqrt{\epsilon_Y} \left( \alpha \ket*{-\frac{1}{2}} +\beta \mathrm{e}^{\mathrm{i}\phi} \ket*{+\frac{1}{2}} \right) + \mathrm{\, higher \, order}
\label{decoded-state}
\end{eqnarray}
Following this decoding, a conditional excitation and measurement of the ancillary electron spin reveals whether the state is corrupted, and if so, the nature of the error. For example, a conditional swap from the nuclear $\ket{-1/2}, \ket{+1/2}$ subspace onto the electron ancilla generates a full electron-nuclear state of the form
\begin{equation}
\ket{\Psi} \sim 
\mathrm{(remaining \, terms)} \ket{0_\mathrm{ancilla}} +
\sqrt{\epsilon_Y} \left( \alpha \ket*{-\frac{1}{2}} +\beta \mathrm{e}^{\mathrm{i}\phi} \ket*{+\frac{1}{2}} \right) \ket{1_\mathrm{ancilla}} 
\end{equation} 
If a subsequent projective measurement of the electron spin ancilla yields $\ket{1}$, we conclude that there was a $S_Y$ error. If it yields $\ket{0}$, we can iterate as shown in Fig.~\ref{schematic} until the ancilla measurement yields $\ket{1}$. In this way we can identify the error cases (i.e., which of $I,S_Z,S_X$, or $S_Y$ occurred), the individual terms of Eqn.~\ref{decoded-state}, and thus the original state $\ket{\psi}$.

\begin{figure}
\includegraphics[width=16cm]{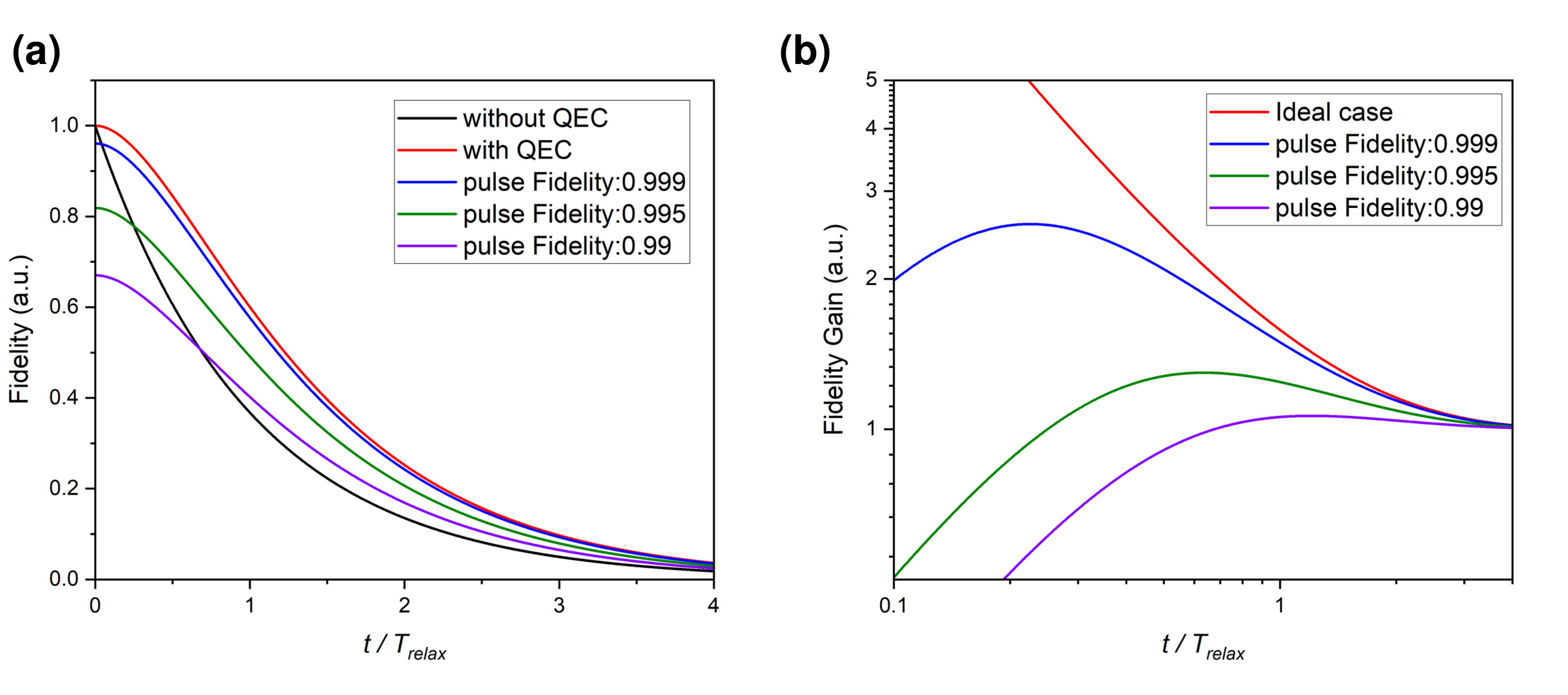}
\caption{ (a) The qubit fidelity as a function of time without (black) and with (red) error correction. The detrimental effect of imperfect magnetic resonance pulses is simulated for pulse fidelities of 0.999 (blue) , 0.995 (green), and 0.99 case (purple). (b) Fidelity gain (defined as the ratio of the infidelities of the uncorrected and corrected cases) for a range of pulse fidelities.}
\label{fidelity}
\end{figure}

The algorithm fails if the measured state does not yield one of these four ($I,S_Z,S_X,S_Y$) outcomes; this corresponds to a higher-order error case and has probability of order $(t/T_\mathrm{relax})^2$. Therefore, as long as $t$ is short compared to $T_\mathrm{relax}$ we obtain a fidelity gain corresponding to the first-order error probability. This is shown in Fig~\ref{fidelity}, which shows how the fidelity varies with $t/T_\mathrm{relax}$ under the assumption that the relaxation rate is isotropic (i.e., rates for $X,Y$ or $Z$ errors are equal) with (orange) and without (blue) error correction. 

The operations implementing this protocol on an electronic $S=1/2$ qubit coupled to a nuclear spin qudit $I=7/2$ correspond to simple microwave or radio-frequency pulses, as long as the spin Hamiltonian includes a Zeeman term (provided by an external magnetic field), a hyperfine coupling between the electronic and nuclear spins, and a term (such as a nuclear quadrupole interaction) lifting the degeneracy of transitions within the nuclear manifold. In a realistic apparatus encoding and decoding pulse sequences have durations of the order of tens of microseconds ~\cite{atzori2021radiofrequency, hussain2018coherent, george2010electron, golter2014optically}. For a condensed matter spin system with relaxation times of the order of tens of milliseconds (as has been reported in various isotopically enriched materials ~\cite{balasubramanian2009ultralong, tyryshkin2012electron}), $t/T_\mathrm{relax} \sim 10^3$ and our scheme extends the qubit coherence by multiple orders of magnitude. 

Given that the entire encoding and decoding procedure requires about 40 pulses,  the effectiveness of the protocol is naturally sensitive to the fidelity of the individual spin manipulation operations. Fig~\ref{fidelity} shows how imperfect magnetic resonance pulses compromise the overall fidelity; individual pulse fidelities must exceed about 0.99 for the protocol to offer any advantage over natural relaxation. 

In particular physical systems the range of available operations is wider, and this can be exploited to implement the protocol more efficiently with fewer operations. For example, through electric field induced modulation of the nuclear quadrupole interaction of $^{123}$Sb nuclear spins in Si, it is possible to drive resonant $\Delta m=\pm 2$ transitions within the $I=7/2$ nuclear spin manifold~\cite{asaad2020coherent}; this significantly reduces the number of SWAP operations ($\pi$-pulses) required in the encoding and decoding sequences and therefore relaxes the threshold for the pulse fidelity to 0.98.

Interestingly, the effectiveness of our protocol corresponds to what is possible in distance-3 codes in qubit error correction. (Distance-3 codes fail when simultaneous errors occur on more than two qubits; for a single-qubit error rate $1/T_\mathrm{relax}$, the probability of two or more simultaneous errors in time $t$ is $\sim (t/T_\mathrm{relax})^2$, as we find above for our protocol.) Based on the quantum Hamming bound ~\cite{gottesman1996class}, the minimal number of qubits for building a distance-3 error correction code is 5, requiring a Hilbert space dimension of 32. Thus, our spin-7/2 qudit-based code has advantages over the existing qubit counterpart, in both the size of Hilbert space and the number of actual physical objects over which quantum control is required. We can generalise this resource argument: for a qubit-based $[n,k,2t+1]$ code,  the quantum Hamming bound 
\begin{equation}
2^k \sum^t_{l=0} 3^l 
\left( \begin{smallmatrix} n \\ l \end{smallmatrix} \right)
\leq 2^n
\end{equation}
determines the number of qubits required~\cite{gottesman1996class}. Thus a distance-5 encoding of a single logical qubit requires 11 physical qubits, i.e., a Hilbert space of dimension 2048. 

The strategy for fault-tolerant encoding on spin qudits can be generalized as follows. Inspired by previous studies and the basic concept of the original GKP code, we suggest code-words comprising $\sim N$ periodic terms on a $\sim 4N^2 $-dimensional qudit. The logical code-words supporting $N^{th}$ order error correction can be generated in the form
\begin{equation}
\label{codeN}
\begin{aligned}
\ket{0_L} &= a_{0}\ket{-S} + \sum^{N}_{i=1}a_{i}\ket{-S+(Ai-B)}\\
\ket{1_L} &= b_{0}\ket{+S} + \sum^{N}_{i=1}b_{i}\ket{+S-(Ai-B)},\\
\end{aligned}
\end{equation}
where the parameters $S, A, B, a_i, b_i$ are chosen to satisfy the Knill-Laflamme criteria (see Supplementary Information). Here we present example cases of solutions corresponding to $d = 4N(N+1)$ and $d = 4N(N+1)+2$, as shown in Table \ref{codewords}.

\begin{table}[htp]

\caption{Example code-words for $N^\mathrm{th}$-order error correction on high-spin qudits.}
\begin{center}
\begin{tabular}{|c|c|c|}

\hline
$N^\mathrm{th}$ order, $d$-dim & $\ket*{0_L}$ Representation in $z$-basis & $\ket*{1_L}$ Representation in $z$-basis \\ \hline\hline
$N=1, d=8$ (spin 7/2) & 
$+\sqrt{\frac{3}{10}} \ket*{-\frac{7}{2}} + \sqrt{\frac{7}{10}}\ket*{+\frac{3}{2}}$ & $-\sqrt{\frac{3}{10}} \ket*{+\frac{7}{2}} + \sqrt{\frac{7}{10}}\ket*{-\frac{3}{2}}$ \\ \hline
$N=1, d=10$ (spin 9/2) & 
$+\sqrt{\frac{1}{4}} \ket*{-\frac{9}{2}} + \sqrt{\frac{3}{4}}\ket*{+\frac{3}{2}}$ & $+\sqrt{\frac{1}{4}} \ket*{+\frac{9}{2}} + \sqrt{\frac{3}{4}}\ket*{-\frac{3}{2}}$\\ \hline
$N=2, d=24$ (spin 23/2) & 
$\makecell{ +\sqrt{\frac{125}{1482}} \ket*{-\frac{23}{2}} + \sqrt{\frac{874}{1482}}\ket*{-\frac{5}{2}} \\ +\sqrt{\frac{483}{1482}} \ket*{+\frac{15}{2}} } $ & $ \makecell{ -\sqrt{\frac{125}{1482}} \ket*{+\frac{23}{2}} + \sqrt{\frac{874}{1482}}\ket*{+\frac{5}{2}} \\ +\sqrt{\frac{483}{1482}} \ket*{-\frac{15}{2}} } $  
\\ \hline
$N=2, d=26$ (spin 25/2) & $ \makecell{ +\sqrt{\frac{1}{16}} \ket*{-\frac{25}{2}}  +  \sqrt{\frac{10}{16}}\ket*{-\frac{5}{2}} \\ +\sqrt{\frac{5}{16}} \ket*{+\frac{15}{2}}}  $ & $ \makecell{ +\sqrt{\frac{1}{16}} \ket*{+\frac{25}{2}} + \sqrt{\frac{10}{16}}\ket*{+\frac{5}{2}} \\ +\sqrt{\frac{5}{16}} \ket*{-\frac{15}{2}}} $\\ \hline
$N=3, d=50 $  (spin 49/2) & 
 $ \makecell{+\sqrt{\frac{1}{64}} \ket*{-\frac{49}{2}} + \sqrt{\frac{21}{64}}\ket*{-\frac{21}{2}} \\ + \sqrt{\frac{35}{64}} \ket*{+\frac{7}{2}} +\sqrt{\frac{7}{64}} \ket*{+\frac{35}{2}} }  $ &  $ \makecell{+\sqrt{\frac{1}{64}} \ket*{+\frac{49}{2}} + \sqrt{\frac{21}{64}}\ket*{+\frac{21}{2}} \\ + \sqrt{\frac{35}{64}} \ket*{-\frac{7}{2}} +\sqrt{\frac{7}{64}} \ket*{-\frac{35}{2}} }  $ \\ \hline
\end{tabular}
\end{center}
\label{codewords}
\end{table}

The example code-word in Table~\ref{codewords} shows that distance-5 encoding for a spin qudit is possible with a system of only $d=24$. These codes can offer orders of magnitude improved resource efficiency compared to qubit-based encodings. 

A previous Bosonic quantum error correction proposal exploiting an analogy with generalized Pauli matrix formalism was suggested in Ref.~\cite{Gottesman-arXiv1997}, but this sequence is somewhat different from the perspective of required resources. In the GKP code formalism and its expansion to harmonic oscillator systems, each sign of amplitude damping ($\pm X$) and phase damping ($\pm Z$) is mapped into a different basis~\cite{bartlett2002quantum}. Error correction up to $N$th order requires a Hilbert space dimension of $2 \times (2N+1) \times (2N+1)$, resulting in minimum $d=18$ basis states for first order correction, and $d=50$ for up to second order correction, about twice the resource compared to our proposal here.

We also note that a Hilbert space comprising multiple qudits can also provide a viable code space (for example, encoding for 2$^\mathrm{nd}$ order error correction using  four spin-7/2 qudits; see Supplementary Information), and can offer optimized codewords when the size or number of spins are limited in specific experimental implementations.

Finally, we address the comparison between $T_1$ (bit error) and $T_2$ (phase error) relaxation in spin systems. Often in the solid state $T_2$ is found to be much shorter than $T_1$; practical proposals have therefore concentrated on phase error protection~\cite{chiesa2020molecular} for pragmatic reasons. However, it has been found in certain systems that careful optimisation of the environment can significantly reduce phase relaxation, for example by isotopically purifying the $^{28}$Si or $^{12}$C host environments for P donors~\cite{tyryshkin2012electron,tyryshkin2003electron} or NV centres~\cite{balasubramanian2009ultralong} respectively, or by engineering the ligand environment in molecular magnets~\cite{Wedge2012}. When phase relaxation mechanisms are suppressed, $T_1$ relaxation will become increasingly important, and error correction algorithms addressing both error classes become important. Furthermore, the strategy that we present here allows for the possibility of checking for each error independently. Under the circumstances that $T_2 << T_1$ and projective measurement of the ancilla electron spin is expensive, an optimal practical implementation might check more frequently for $S_Z$ errors than for $S_X$ and $S_Y$ errors.

\begin{acknowledgments}
This project was supported by the European Union's Horizon 2020 research and innovation programme under grant agreements 862893 (FATMOLS) and 863098 (SPRING).
\end{acknowledgments}

%\bibliography{faulttoleranttrunc}

\newpage
\newpage

\setcounter{page}{1}
\setcounter{equation}{0}

\begin{center}
{\large\bf{Supplementary Information: Fault-tolerant qubit encoding using a spin-7/2 qudit}}

Sumin Lim, Junjie Liu, and Arzhang Ardavan

\end{center}

\section{Candidates for logical qubit $\ket{0_L}$  and $\ket{1_L}$ states}
\subsection{Encoding for first-order error correction} \label{first-order-7halves}

In order to enable first order error correction for $S_X$, $S_Y$ and $S_Z$, the logical qubits, $\ket{0_L}$ and $\ket{1_L}$, should satisfy the Knill-Laflamme criteria
\begin{equation}
\label{KLcriteria}
\begin{aligned}
\bra{0_L} S_i^l S_j^k \ket{1_L}  &=  0,\\
\bra{0_L} S_i^l S_j^k \ket{0_L}   &=  \bra{1_L} S_i^l S_j^k \ket{1_L}  = \delta_{i,j}\delta_{l,k},
\end{aligned}
\end{equation}
where $i, j = X, Y$ or $Z$ and $l, k = 0$ or $1$.  It is not strictly necessary for the second equation to be identical to the Kronecker delta. However, it (a) helps to distinguish the corrupted state from pure states and (b) does not change the minimum hardware required for the quantum error correction (QEC).  
%\footnote
({While it is not necessary for all error states to be orthogonal to each other, they need to be linearly independent to ensure that different errors can be distinguished. Hence, lifting the orthogonality condition implied by the Kronecker delta would not change the dimension of the vector space spanned by the logic qubits and all possible error states.}) 
Effectively, Eqn.~\ref{KLcriteria} requires the following 8 states:
\begin{equation}
\label{states}
\ket{0_L},  S_X\ket{0_L}, S_Y\ket{0_L}, S_Z\ket{0_L}, \ket{1_L}, S_X\ket{1_L}, S_y\ket{1_L} , S_z\ket{1_L}
\end{equation}
 to be non-zero and mutually orthogonal. The smallest Hilbert space that can accommodate these states is 8 dimensional, in which case the 8 states in Eqn.~\ref{states} form a complete orthogonal basis set.  Hence, the minimum spin qudit that can potentially allow corrections for all first order errors is a $S = 7/2$ system. 

Finding all possible solutions for Eqn.~\ref{KLcriteria} with $S = 7/2$ is cumbersome as it involves solving multiple coupled second-order polynomials. Thus, we confirmed the existence of $\ket{0_L}$ and $\ket{1_L}$ combinations by finding examples satisfying Eqn.~\ref{KLcriteria} numerically. The simplest solution we found is 
\begin{equation}
\label{code1}
\begin{aligned}
\ket{0_L} &= \sqrt{\frac{3}{10}} \ket*{-\frac{7}{2}} + \sqrt{\frac{7}{10}}\ket*{+\frac{3}{2}}\\
\ket{1_L} &= -\sqrt{\frac{7}{10}} \ket*{-\frac{3}{2}} + \sqrt{\frac{3}{10}}\ket*{+\frac{7}{2}}.
\end{aligned}
\end{equation}
It is straightforward to verify that these logical qubits meet the criteria listed in Eqn.~\ref{KLcriteria}. It is worth pointing out, however, that this solution (which was previously identified by Gross~[Phys.\ Rev.\ Lett.\ {\bf 127}, 010504 (2021)]) is not unique. For example, the following states:
\begin{equation}
\label{code2}
\begin{aligned}
\ket{0_L} &= -\sqrt{\frac{21}{64}} \ket*{-\frac{5}{2}} + \sqrt{\frac{21}{64}}\ket*{-\frac{1}{2}}+\sqrt{\frac{7}{64}}\ket*{+\frac{3}{2}}+\sqrt{\frac{15}{64}}\ket*{+\frac{7}{2}}\\
\ket{1_L} &= -\sqrt{\frac{15}{64}} \ket*{-\frac{7}{2}} - \sqrt{\frac{7}{64}}\ket*{-\frac{3}{2}}-\sqrt{\frac{21}{64}}\ket*{+\frac{1}{2}}+\sqrt{\frac{21}{64}}\ket*{+\frac{5}{2}}
\end{aligned}
\end{equation}
can also be used for correcting all first order errors. However, the encoding and decoding pulse sequences for Eqn.~\ref{code2}, when implemented on a spin qudit, are significantly more complicated than those for Eqn.~\ref{code1}; thus we choose to use Eqn.~\ref{code1} as the working example for all discussions presented in this work.

For qudits with $S > 7/2$, similar codewords for first order QEC can be found easily. For instance, in $S = 9/2$ systems (e.g.\ the nuclear spins of Bi and Sr) the simplest logical qubits are
\begin{equation}
\label{code_highS}
\begin{aligned}
\ket{0_L} &= -\sqrt{\frac{1}{4}} \ket*{-\frac{9}{2}} + \sqrt{\frac{3}{4}}\ket*{+\frac{3}{2}}\\
\ket{1_L} &= +\sqrt{\frac{3}{4}} \ket*{-\frac{3}{2}} + \sqrt{\frac{1}{4}}\ket*{+\frac{9}{2}}.
\end{aligned}
\end{equation}

\subsection{Encoding for second and higher-order error correction}

One can apply the same strategy to develop an efficient protocol for correcting high-order quantum errors using  qudits. To correct an $N^\mathrm{th}$-order error, the logical qubits should meet the following criteria:
\begin{equation}
\label{KLcriteria_highorder}
\begin{aligned}
\bra{0_L} S_i^l S_j^k \ket{1_L}  &=  0,\\
\bra{0_L} S_i^l S_j^k \ket{0_L}   &=  \bra{1_L} S_i^l S_j^k \ket{1_L},
\end{aligned}
\end{equation}
where $i, j = X, Y$ or $Z$ and $l, k = 0$, $1 \ldots N$.

Here, we propose a general approach for designing the logical qubits within  $4N(N+1)$ or $4N(N+1)+2$ dimensional spaces, hence enabling any $N^\mathrm{th}$-order error correction to be implemented with a $S = [4N(N+1) - 1]/2$ or a $S = [4N(N+1) + 1]/2$ qudit, respectively. 

In the case of $S = [4N(N+1) + 1]/2$, both $\ket{0_L}$ and $\ket{1_L}$ are superpositions of the $N+1$ $S_z$ eigenstates such that
\begin{equation}
\label{codeN}
\begin{aligned}
\ket{0_L} &= a_{0}\ket{-S} + \sum^{N}_{i=1}a_{i}\ket{-S+(Ai-B)}\\
\ket{1_L} &= b_{0}\ket{+S} + \sum^{N}_{i=1}b_{i}\ket{+S-(Ai-B)},\\
\end{aligned}
\end{equation}
where $A = 4N+2$, $B = 0$, and the coefficients for $\ket{0_L}$,  $\{a_i\}$ ($i = 0$, $1 \ldots N$),  are calculated by solving the following equations:
\begin{equation}
\label{coeff_a}
\begin{aligned}
%\bra{0_L}\ket{0_L}=1 ,\\
\langle 0_L \vert 0_L \rangle = 1 ,\\
\bra{0_L} S_{X}^j S_{Z} S_{X}^j \ket{0_L}=0 ,
\end{aligned}
\end{equation}
where $ 0\le j \le N-1 $. Eqn.~\ref{coeff_a} imposes $N + 1$ constraints on $\{a_i\}$. Indeed, Eqn.~\ref{coeff_a} can be rewritten as $N+1$ real linear equations with $\{|a_i|^2\}$ being $N+1$ independent variables. Solving them gives a set of positive $|a_i|^2$ and we choose the root $a_i = |a_i|$ for constructing $\ket{0_L}$. $\ket{1_L}$ can then be defined with $b_0 = a_0$ and $b_{i} = a_{i}$ for $ 1\le i \le N$. 

In the case of $S = [4N(N+1) - 1]/2$, both $\ket{0_L}$ and $\ket{1_L}$ are constructed in similar form, but with $A = 4N+2$ and $B=1$. The coefficients are obtained with almost the same conditions, except for $b_0$, which is defined by $b_0 = -a_0$.

{We do not offer a proof that this approach works in all cases, though we find that it works for a wide range of cases that we have explored, and we have found none where it does not.}

By way of illustration, the approach yields logical qubits for 3rd-order QEC ($N = 3$) on a a $S = 47/2$ system, given by
\begin{equation}
\label{code47}
\begin{aligned}
\ket{0_L} &= +\sqrt{\frac{16807}{796302}} \ket*{-\frac{47}{2}} + \sqrt{\frac{260145}{796302}}\ket*{-\frac{21}{2}}+ \sqrt{\frac{425867}{796302}}\ket*{+\frac{7}{2}}+\sqrt{\frac{93483}{796302}}\ket*{+\frac{35}{2}}\\
\ket{1_L} &= -\sqrt{\frac{16807}{796302}} \ket*{+\frac{47}{2}} + \sqrt{\frac{260145}{796302}}\ket*{+\frac{21}{2}}+ \sqrt{\frac{425867}{796302}}\ket*{-\frac{7}{2}}+\sqrt{\frac{93483}{796302}}\ket*{-\frac{35}{2}},
\end{aligned}
\end{equation}
and 4th-order QEC ($N = 4$) on a $S=81/2$ system, given by
\begin{equation}
\label{code81}
\begin{aligned}
\ket{0_L} &= +\sqrt{\frac{1}{256}} \ket*{-\frac{81}{2}} + \sqrt{\frac{36}{256}}\ket*{-\frac{45}{2}}+ \sqrt{\frac{126}{256}}\ket*{-\frac{9}{2}}+\sqrt{\frac{84}{256}}\ket*{+\frac{27}{2}}+\sqrt{\frac{9}{256}}\ket*{+\frac{63}{2}}\\
\ket{1_L} &= +\sqrt{\frac{1}{256}} \ket*{+\frac{81}{2}} + \sqrt{\frac{36}{256}}\ket*{+\frac{45}{2}}+ \sqrt{\frac{126}{256}}\ket*{+\frac{9}{2}}+\sqrt{\frac{84}{256}}\ket*{-\frac{27}{2}}+\sqrt{\frac{9}{256}}\ket*{-\frac{63}{2}},
\end{aligned}
\end{equation}
which satisfy all KL criteria in Eqn.~\ref{KLcriteria_highorder}.

Finally, the spin operator identities $\sum_{i=X,Y,Z} S_i^2=S(S+1)$ and $[S_i, S_j]={\mathrm i} S_k$ encapsulate the behaviour of the physical system that embodies the quantum information and is subject to noise. We note that they constrain the evolution of second-order errors in such a way that it may be possible to perform high-order QEC ($N \ge 2$) with a qudit smaller than $S = [4N(N+1) - 1]/2$.

\subsection{Encoding for multiple qudits} 

The same approach may be used to build logical qubits for a physical system comprising multiple qudits. For example, for the three spin-3/2 qudits, (labeled as A, B, and C) the codewords given by 
\begin{equation}
\label{codeM1}
\begin{aligned}
\ket{0_L} &= +\sqrt{\frac{1}{4}} \ket*{-\frac{3}{2}}_{A,B,C} + \sqrt{\frac{3}{4}}\ket*{+\frac{1}{2}}_{A,B,C}\\
\ket{1_L} &= +\sqrt{\frac{1}{4}} \ket*{+\frac{3}{2}}_{A,B,C} + \sqrt{\frac{3}{4}}\ket*{-\frac{1}{2}}_{A,B,C}
\end{aligned}
\end{equation}
satisfy all KL criteria in Eqn.~\ref{KLcriteria_highorder} up to first order.
And for four spin-7/2 qudits, (labeled as A, B, C, and D), the codewords given by  
\begin{equation}
\label{codeM2}
\begin{aligned}
\ket{0_L} &= +\sqrt{\frac{2}{16}} \ket*{-\frac{7}{2}}_{A,B,C,D} +\sqrt{\frac{7}{16}} \ket*{-\frac{3}{2}}_{A,B,C,D}+\sqrt{\frac{7}{16}} \ket*{+\frac{5}{2}}_{A,B,C,D}\\
\ket{1_L} &= +\sqrt{\frac{2}{16}} \ket*{+\frac{7}{2}}_{A,B,C,D} +\sqrt{\frac{7}{16}} \ket*{+\frac{3}{2}}_{A,B,C,D} -\sqrt{\frac{7}{16}} \ket*{-\frac{5}{2}}_{A,B,C,D},
\end{aligned}
\end{equation}
satisfy all KL criteria in Eqn.~\ref{KLcriteria_highorder} up to second order.

\section{Pulse sequences for encoding and decoding logical qubits}
Here we discuss the physical operations that implement the spin-7/2 first order protection presented in the main manuscript and in Section~\ref{first-order-7halves} of this Supplementary Information.
\subsection{Encoding pulse sequence}
Without loss of generality, we may assume a qubit $\ket{\psi}$ is initially stored in a superposition between $\ket{-7/2}$ and $\ket{-5/2}$ states, such that $\ket{\psi} = \alpha\ket{-7/2}+\beta e^{{\mathrm i}\theta}\ket{-5/2}$, where $\alpha$, $\beta$ and $\theta$ encode the quantum information.  In the $S_Z$ basis, 
\begin{equation}
\ket{\psi} = \left[
\begin{matrix}
\alpha, & \beta e^{{\mathrm i}\theta}, & 0,  & 0, & 0, & 0, & 0, & 0 \\
\end{matrix}
\right]^T.
\end{equation}
The first four pulses in the encoding sequence [Fig. 1(c) in the main text], i.e.\ the first $\pi$-pulse, $U_{\theta1}$, $U_{\theta2}$ and the second $\pi$-pulse, tranform $\ket{\psi}$ into
\begin{equation}
\left[
\begin{matrix}
\sqrt{\frac{3}{10}}\alpha, & -\sqrt{\frac{7}{10}}\beta e^{i\theta}, & \sqrt{\frac{7}{10}}\alpha,  & \sqrt{\frac{3}{10}}\beta e^{i\theta}, & 0, & 0, & 0, & 0\\
\end{matrix}
\right]^T.
\end{equation}
The subsequent 8 $\pi$-pulses further transform the state into the encoded state
\begin{equation}
\label{state_encode}
\ket{\Psi^{\mathrm{enc}}} = \left[
\begin{matrix}
\sqrt{\frac{3}{10}}\alpha, & 0, & -\sqrt{\frac{7}{10}}\beta e^{i\theta},  & 0, & 0, & \sqrt{\frac{7}{10}}\alpha, & 0, & \sqrt{\frac{3}{10}}\beta e^{i\theta}\\
\end{matrix}
\right]^T = \alpha\ket{0_L} + \beta e^{i\theta}\ket{1_L}.
\end{equation}
As shown in Eqn.~\ref{state_encode}, the entire encoding sequence maps the original qubit onto the corresponding superposition of QEC logical qubit states $\ket{0_L}$ and $\ket{1_L}$.

\subsection{Decoding pulse sequence}
After the storage period during which first-order errors may occur, the density matrix of the nuclear spin can be written as 
\begin{equation}
\label{state_error}
\rho(t) = (1-\epsilon)I  + \epsilon_X S_X \rho(0) S_X^\dagger + \epsilon_Y S_Y \rho(0) S_Y^\dagger + \epsilon_Z S_Z \rho(0) S_Z^\dagger + O(\epsilon^2 ),
\end{equation}
where $\epsilon_i \sim t/T_{\mathrm{relax},i}$ is indicative of the scale of the error. Interpreting the decoding sequence using the general form of Eqn.~\ref{state_error} can be laborious. However, as we discussed in the previous section,  the logical qubit states and their first-order errors form an orthogonal basis for the Hilbert state (Eqn.~\ref{states}). Therefore,  we demonstrate the decoding sequence for the states $\ket{\Psi^{\mathrm{enc}}}$,  $S_X\ket{\Psi^{\mathrm{enc}}}$, $S_X\ket{\Psi^{\mathrm{enc}}}$ and $S_Y\ket{\Psi^{\mathrm{enc}}}$ individually. As we will show in the following section, the decoding sequence converts the states in Eqn.~\ref{states} into another set of orthogonal states that are spectroscopically distinguishable, allowing errors to be identified using the corresponding projective measurements. The general error form is the linear combination of these decoupled orthogonal states.

\subsubsection{Error free}
If no error occurs,  the spin remains in the state $\ket{\Psi^{\mathrm{enc}}}$ state.  To decode this state, the first seven successive $\pi$-pulses in the decoding sequence $U_\mathrm{dec}$ transform $\ket{\Psi^{\mathrm{enc}}}$ into
\begin{equation}
\left[
\begin{matrix}
\sqrt{\frac{3}{10}}\alpha, & \sqrt{\frac{7}{10}}\alpha, & 0,  & 0, & 0, & 0, & -\sqrt{\frac{7}{10}}\beta e^{i\theta}, & \sqrt{\frac{3}{10}}\beta e^{i\theta}\\
\end{matrix}
\right]^T,
\end{equation}
and the subsequent two simultaneous $U_{- \theta 1}$-pulses further rotate it to the state
\begin{equation}
\left[
\begin{matrix}
\alpha, & 0, & 0, & 0, & 0, & 0, & 0, & \beta e^{i\theta}\\
\end{matrix}
\right]^T.
\end{equation}
The remaining pulses do not act on either $\ket{-7/2}$ or $\ket{7/2}$, so this is the final decoded state.

\subsubsection{$S_{Z}$ error}
An $S_Z$ error will lead to a non-zero component for the state
\begin{equation}
S_Z\ket{\Psi^{\mathrm{enc}}} = \left[
\begin{matrix}
-\sqrt{\frac{7}{10}}\alpha, & 0, & \sqrt{\frac{3}{10}}\beta e^{i\theta},  & 0, & 0, & \sqrt{\frac{3}{10}}\alpha, & 0, & \sqrt{\frac{7}{10}}\beta e^{i\theta}\\
\end{matrix}
\right]^T.
\end{equation}
To decode this component,  the first seven $\pi$-pulses in $U_\mathrm{dec}$ transform the state to
\begin{equation}
\left[
\begin{matrix}
-\sqrt{\frac{7}{10}}\alpha, & \sqrt{\frac{3}{10}}\alpha, & 0,  & 0, & 0, & 0, & \sqrt{\frac{3}{10}}\beta e^{i\theta}, & \sqrt{\frac{7}{10}}\beta e^{i\theta}\\
\end{matrix}
\right]^T,
\end{equation}
and the following two $U_{- \theta 1}$-pulses transform it to
\begin{equation}
\left[
\begin{matrix}
0, & \alpha, & 0,  & 0, & 0, & 0, & \beta e^{i\theta}, & 0\\
\end{matrix}
\right]^T,
\end{equation}
at which point the decoding process is complete. Any $S_Z$ error is readily detectable by a subsequent projective measurement on the on the $m_I = \pm 5/2$ hyperfine transition of the electron spin ancilla.

\subsubsection{$S_{X}$ error}
A $S_X$ error will lead to a non-zero component for the state
\begin{equation}
\begin{aligned}
S_X&\ket{\Psi^{\mathrm{enc}}}\\
&= \left[
\begin{matrix}
0, & +\sqrt{\frac{1}{10}}\alpha -\sqrt{\frac{4}{10}}\beta e^{i\theta}, & 0,  & -\sqrt{\frac{5}{10}}\beta e^{i\theta}, & \sqrt{\frac{5}{10}}\alpha, & 0, & \sqrt{\frac{4}{10}}\alpha+\sqrt{\frac{1}{10}}\beta e^{i\theta}, & 0 \\
\end{matrix}
\right]^T.
\end{aligned}
\end{equation}
To decode this component,  the first seven $\pi$-pulses in $U_\mathrm{dec}$ transform the state to
\begin{equation}
\left[
\begin{matrix}
0, & 0, & +\sqrt{\frac{1}{10}}\alpha -\sqrt{\frac{4}{10}}\beta e^{i\theta},  & +\sqrt{\frac{5}{10}}\beta e^{i\theta}, & -\sqrt{\frac{5}{10}}\alpha, & \sqrt{\frac{4}{10}}\alpha+\sqrt{\frac{1}{10}}\beta e^{i\theta}, & 0, & 0 \\
\end{matrix}
\right]^T.
\end{equation}
The two following $U_{- \theta 1}$-pulses do not alter the state since they only act on the \{$\ket{-7/2}$, $\ket{-5/2}$, $\ket{5/2}$ and $\ket{7/2}$\} subspace.  Instead, the two $\pi$-pulses and  $U_{- \theta 3}$ subsequently transform the state to
\begin{equation}
\left[
\begin{matrix}
0, & 0, & -\sqrt{\frac{5}{10}}\beta e^{i\theta},  & +\sqrt{\frac{5}{10}}\alpha, & \sqrt{\frac{5}{10}}\beta e^{i\theta}, & -\sqrt{\frac{5}{10}}\alpha & 0, & 0 \\
\end{matrix}
\right]^T,
\end{equation}
which is further transformed by the subsequent five $\pi$-pulses into
\begin{equation}
\left[
\begin{matrix}
0, & 0, & \sqrt{\frac{5}{10}}\alpha,  & -\sqrt{\frac{5}{10}}\alpha, & -\sqrt{\frac{5}{10}}\beta e^{i\theta}, & -\sqrt{\frac{5}{10}}\beta e^{i\theta}, & 0, & 0 \\
\end{matrix}
\right]^T.
\end{equation}
The final part of the decoding sequence, consisting of two $U_{ \theta 4}$-pulses and two $\pi$-pulses, transforms the state into the form
\begin{equation}
\left[
\begin{matrix}
0, & 0, & \alpha,  & 0, & 0, & \beta e^{i\theta}, & 0, & 0 \\
\end{matrix}
\right]^T,
\end{equation}
which allows any $S_X$ error to be detected via a measurement on the $m_I = \pm 3/2$ hyperfine transition of the electron ancilla.

\subsubsection{$S_{Y}$ error}
The underlying principle for decoding $S_Y$ error is equivalent to that for $S_X$ error, albeit the exact states involved are slightly different. A $S_Y$ error will lead to the state
\begin{equation}
\begin{aligned}
S_Y&\ket{\Psi^{\mathrm{enc}}}\\
&=\left[
\begin{matrix}
0, & -\sqrt{\frac{1}{10}}\alpha -\sqrt{\frac{4}{10}}\beta e^{i\theta}, & 0,  & \sqrt{\frac{5}{10}}\beta e^{i\theta}, & \sqrt{\frac{5}{10}}\alpha, & 0, & -\sqrt{\frac{4}{10}}\alpha+\sqrt{\frac{1}{10}}\beta e^{i\theta}, & 0 \\
\end{matrix}
\right]^T.
\end{aligned}
\end{equation}
The first seven $\pi$-pulses in $U_\mathrm{dec}$ transform the state to
\begin{equation}
\left[
\begin{matrix}
0, & 0, & -\sqrt{\frac{1}{10}}\alpha -\sqrt{\frac{4}{10}}\beta e^{i\theta},  & -\sqrt{\frac{5}{10}}\beta e^{i\theta}, & -\sqrt{\frac{5}{10}}\alpha, & \sqrt{\frac{4}{10}}\alpha-\sqrt{\frac{1}{10}}\beta e^{i\theta}, & 0, & 0 \\
\end{matrix}
\right]^T.
\end{equation}
The following two $U_{- \theta 1}$-pulses, the two $\pi$-pulses and $U_{- \theta 3}$ transform it to
\begin{equation}
\left[
\begin{matrix}
0, & 0, & +\sqrt{\frac{5}{10}}\beta e^{i\theta},  & -\sqrt{\frac{5}{10}}\alpha, & +\sqrt{\frac{5}{10}}\beta e^{i\theta}, & -\sqrt{\frac{5}{10}}\alpha & 0, & 0 \\
\end{matrix}
\right]^T.
\end{equation}
Then, the five $\pi$-pulses reorder the state to 
\begin{equation}
\left[
\begin{matrix}
0, & 0, & +\sqrt{\frac{5}{10}}\alpha,  & +\sqrt{\frac{5}{10}}\alpha, & -\sqrt{\frac{5}{10}}\beta e^{i\theta}, & +\sqrt{\frac{5}{10}}\beta e^{i\theta}, & 0, & 0 \\
\end{matrix}
\right]^T,
\end{equation}
The final step for decoding is performed by the two $U_{ \theta 4}$-pulses and two  $\pi$-pulses, resulting the final state of
\begin{equation}
\left[
\begin{matrix}
0, & 0, & 0,  & \alpha, & \beta e^{i\theta}, & 0, & 0, & 0 \\
\end{matrix}
\right]^T.
\end{equation}
Again, this state allows the detection for $S_Y$ errors using projective measurements on the $m_I = \pm 1/2$ hyperfine transition of the electron spin ancilla.

\end{document}